\documentclass[reprint,prl,aps]{revtex4-1}
\usepackage[pdftex]{graphicx}
\usepackage{color}
\usepackage{amsmath,amsthm,amssymb}
\usepackage[version=3]{mhchem}

\begin{document}

\title{Accurate barrier heights using diffusion Monte Carlo}
\author{Kittithat Krongchon}
\author{Brian Busemeyer}
\author{Lucas K. Wagner}
\affiliation{Department of Physics, University of Illinois at Urbana-Champaign}
\date{\today}

\begin{abstract}
Fixed node diffusion Monte Carlo (DMC) has been performed on a test set of forward and reverse barrier heights for 19 non-hydrogen-transfer reactions, and the nodal error has been assessed.
The DMC results are robust to changes in the nodal surface, as assessed by using different mean-field techniques to generate single determinant wave functions.
Using these single determinant nodal surfaces, DMC results in errors of 1.5(5) kcal/mol on barrier heights.
Using the large data set of DMC energies, we attempted to find good descriptors of the fixed node error.
It does not correlate with a number of descriptors including change in density, but does correlate with the gap between the highest occupied and lowest unoccupied orbital energies in the mean-field calculation.
\end{abstract}

\maketitle

\section{Introduction}

Quantum Monte Carlo  techniques are promising as a route towards scalable and accurate chemical calculations.
In particular, fixed-node diffusion Monte Carlo (DMC) appears to offer a good compromise between efficiency and accuracy.
This method is particularly attractive because their computational cost scales mildly with system size, ${\cal O}(N_e^{3-4})$, where $N_e$ is the number of electrons in the system, and it does not require basis set extrapolation since it functions in the complete basis set limit~\cite{foulkes_quantum_2001}.
DMC has been applied to both periodic and open boundary Hamiltonians and can obtain accurate results currently up to around 1000 electrons~\cite{wagner_discovering_2016}.
These aspects make DMC an interesting possibility for studying large reactions such as those on surfaces.

DMC essentially finds the lowest energy wave function compatible with the nodes of a trial wave function.
As such, it has an upper bound property to the exact ground state energies, and the nodes of the trial wave function can be variationally optimized by minimizing the final DMC energy.
Due to the high-dimensionality of the $3N_e-1$ dimensional nodes of the trial wave function, this fixed-node error is sometimes difficult to access. However, in many cases, DMC can still achieve quantitative agreement with experiments, even in systems that otherwise difficult to model with parameter-free methods~\cite{wagner_discovering_2016}.

In order to properly evaluate the performance of DMC on chemical systems, extensive benchmarking is necessary.
In the literature, there are benchmarks of the energies of atomic systems~\cite{brown_energies_2007,sarsa_quantum_2008,seth_quantum_2011,seth_quantum_2011,buendia_quantum_2013},
small molecules including transition metals~\cite{wagner_energetics_2007,dubecky_quantum_2013,zen_molecular_2013,dubecky_quantum_2014,yang_how_2015},
and large benchmark studies have been performed on the G1 set of 55 molecules~\cite{grossman_benchmark_2002,nemec_benchmark_2010}.
A review of applications of DMC to chemical systems is available from Austin \textit{et al.}~\cite{austin_quantum_2012}, and a review of applications of DMC to bulk systems is available from Koloren\v c and Mitas~\cite{kolorenc_applications_2011} as well as Wagner and Ceperley~\cite{wagner_discovering_2016}.
Atomic studies find agreement with experiments on order of 0.23~kcal/mol for ionization potentials, on order of 2.3~kcal/mol for electron affinity, and on order 3~kcal/mol for atomization energies.
They also find that DMC recovers 90-95\% of the correlation energy using single Slater--Jastrow trial wave functions, and around 99\% for multi Slater--Jastrow trial wave functions.
Noncovalent interaction energies were found to agree with CCSD(T)/CBS within 0.1 kcal/mol.
In most cases, energy differences roughly within so-called chemical accuracy of 1 kcal/mol are attainable, particularly with multideterminant trial wave functions.

There have been only a few tests of the performance of QMC methods for reaction barriers.
DMC studies of reaction barriers have been calculated for \ce{H + H2} \cite{barnett_h_1985,reynolds_molecular_1986,anderson_note:_2016}, several organic molecules \cite{grossman_high_1997,kollias_quantum_2004,barborini_reaction_2012,fracchia_multi-level_2014,pakhira_quantum_2015}, surface reactions \cite{kanai_toward_2009,hoggan_quantum_2014,wu_hexagonal_2016}, and others \cite{lu_theoretical_2005,fracchia_barrier_2013}.
These have generally found that DMC can get close to or within chemical accuracy for reaction barriers, often improving  on DFT results.
However, most of these studies have only considered a few molecules at a time, and to our knowledge, no large test sets of reaction barriers computed at the same level of accuracy using DMC have yet been conducted.

In this article, we use a high-throughput implementation of DMC to study the database from Peverati and Truhlar \cite{peverati2014quest,zhao2005benchmark}.
The database consists of non-hydrogen-transfer reactions that involves relatively small molecules.
We evaluate several simple strategies for constructing trial wave functions and assess their performance.
We also present a recipe for QMC calculations of reaction barriers that obtains mean absolute errors of reaction barriers of approximately 1.5(5) kcal/mol, close to so-called chemical accuracy.

\section{Method}
We consider the 19 reactions enumerated in Table~\ref{tab:id}.
The barrier heights are calculated by subtracting the DMC total energies of the products and reactants to the transition states.
We employ the clamped ion approximation to the Hamiltonian of the ions.
\begin{table}
\caption{\label{tab:id}
The non-hydrogen-transfer reactions \cite{zhao2005benchmark} and their corresponding IDs.
The transition states are labeled as TS01 to TS19, following Ref~\cite{peverati2014quest}.
When $+$ separates the reactants or products, the energies of each atom or molecule are calculated in separate DFT, HF, or DMC calculations, whereas when $\ldots$ separates the reactants or products, all the reactants or all the products are simulated in the same calculation together although they are actually seperated in physical space.}

\begin{ruledtabular}
\begin{tabular}{rl}
ID & Reaction \\
\colrule
1 & \ce{H + N2O ->[\mathrm{TS01}] OH + N2} \\
2 & \ce{H + FH ->[\mathrm{TS02}] HF + H} \\
3 & \ce{H + ClH ->[\mathrm{TS03}] HCl + H} \\
4 & \ce{H + FCH3 ->[\mathrm{TS04}] HF + CH3} \\
5 & \ce{H + F2 ->[\mathrm{TS05}] HF + F} \\
6 & \ce{CH3 + FCl ->[\mathrm{TS06}] CH3F + Cl} \\
7 & \ce{F- + CH3F ->[\mathrm{TS07}] FCH3 + F-} \\
8 & \ce{F- $\ldots$ CH3F ->[\mathrm{TS07}] FCH3 $\ldots$ F-} \\
9 & \ce{Cl- + CH3Cl ->[\mathrm{TS09}] ClCH3 + Cl-} \\
10 & \ce{Cl- $\ldots$ CH3Cl ->[\mathrm{TS09}] ClCH3 $\ldots$ Cl-} \\
11 & \ce{F- + CH3Cl ->[\mathrm{TS11}] FCH3 + Cl-} \\
12 & \ce{F- $\ldots$ CH3Cl ->[\mathrm{TS11}] FCH3 $\ldots$ Cl-} \\
13 & \ce{OH- + CH3F ->[\mathrm{TS13}] HOCH3 + F-} \\
14 & \ce{OH- $\ldots$ CH3F ->[\mathrm{TS13}] HOCH3 $\ldots$ F-} \\
15 & \ce{H + N2 ->[\mathrm{TS15}] HN2} \\
16 & \ce{H + CO ->[\mathrm{TS16}] HCO} \\
17 & \ce{H + C2H4 ->[\mathrm{TS17}] CH3CH2} \\
18 & \ce{CH3 + C2H4 ->[\mathrm{TS18}] CH3CH2CH2} \\
19 & \ce{HCN ->[\mathrm{TS19}] HNC} \\

\end{tabular}
\end{ruledtabular}
\end{table}
The HF and DFT calculations were performed using the CRYSTAL code \cite{dovesi2005crystal,dovesi2009crystal09}.
The core electrons were removed using pseudopotentials published by Burkatzki, Filippi, and Dolg \cite{burkatzki2007energy,burkatzki2008energy}.
The use of these pseudopotentials has been justified by Nazarov \textit{et al.} \cite{nazarov_benchmarking_2016}.
We used a Gaussian triple-$\zeta$ basis set with polarization.

The set of orbitals produced by each of these methods was used as the foundation for a Slater--Jastrow-type trial wave function for DMC calculations.
This wave function takes the form:
\begin{align}
\Psi(\textbf{R}) &= \text{Det}\left[\phi^\uparrow_i(r^\uparrow_j)\right]\text{Det}\left[\phi^\downarrow_i(r^\downarrow_j)\right]\text{exp}(J),
\end{align}
where $\textbf{R} = \{r_i\}_{i=1}^N$ is the collection of electron coordinates of the $N$-electron system, $\phi$ is the orbital basis, $i$ and $j$ are electron indices, $\uparrow$ and $\downarrow$ indicate spins, and $J$ is the Jastrow factor as defined in Mit{\'a}{\v{s}} and Martin's paper \cite{mitavs1994quantum}.
The Jastrow factor was optimized in a variational Monte Carlo scheme which minimizes the variance of the local energy of the trial wave function.
DMC was then performed on the Slater--Jastrow wave function to find the best estimate of the ground state energy for each system.
Both the variational and the diffusion Monte Carlo calculations are done within the open source code QWalk \cite{wagner2009qwalk}.
Thus, four DMC methods: DMC(PBE), DMC(HF), DMC(PBE0), and DMC(Min) are considered.
The first three represent DMC calculations whose Slater determinant is generated by the method in parentheses.
The DMC(Min) method is formed by taking the minimum DMC energy among the other three.
Due to the variational principle, DMC(Min) should give the closest upper bound to the ground state energy and would be the canonical DMC result for predictions.

\section{Results}
\begin{figure*}
\includegraphics{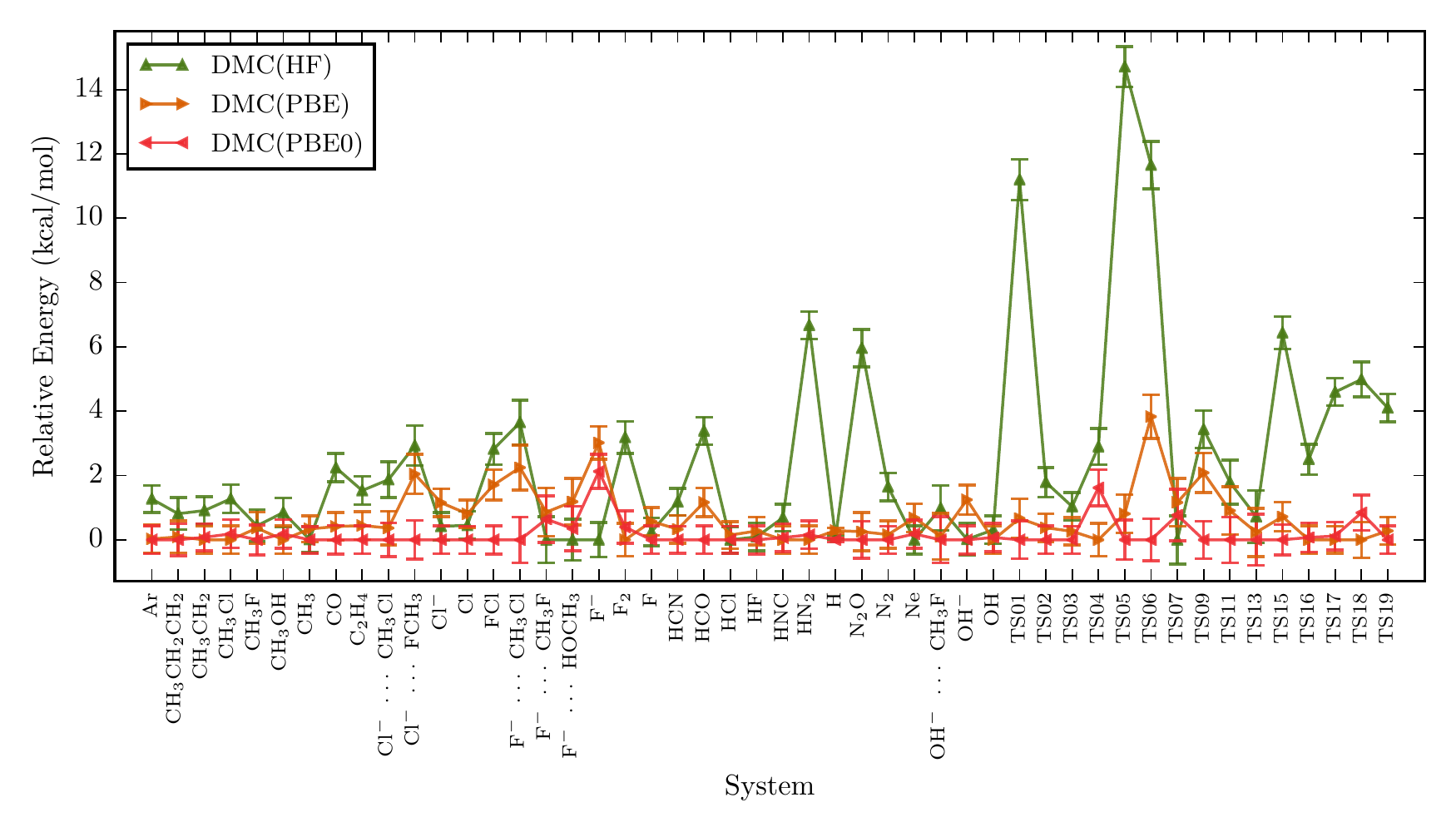}
\caption{\label{fig:edmc}
The DMC total energy relative to the lowest energy among the three functionals.
The error bars are statistical errors from DMC.}
\end{figure*}

The trial wave function performances from each of the DMC approachs are compared in Fig.~\ref{fig:edmc}.
For each system, we show the total DMC energy relative to the lowest energy of the three methods.
The transition states labeled as TS01 to TS19 are defined in Table~\ref{tab:id}.
The \ce{Ne} and \ce{Ar} atoms have also been checked and included in the plot as a comparison to other closed shell systems.
From the plot, the Kohn--Sham orbitals calculated from the PBE0 functional yield the lowest DMC energies for almost all systems studied, except for \ce{F-} and \ce{Ne}, which are closed shell second period atoms, where DMC(HF) outperforms DMC(PBE0).
The fact that HF tends to do better for closed shell second period atoms agrees with the general trend that has been observed, for example, in \ce{C2} and \ce{Si2} \cite{umrigar2007alleviation,rasch2014communication}.

\begin{figure*}
\includegraphics{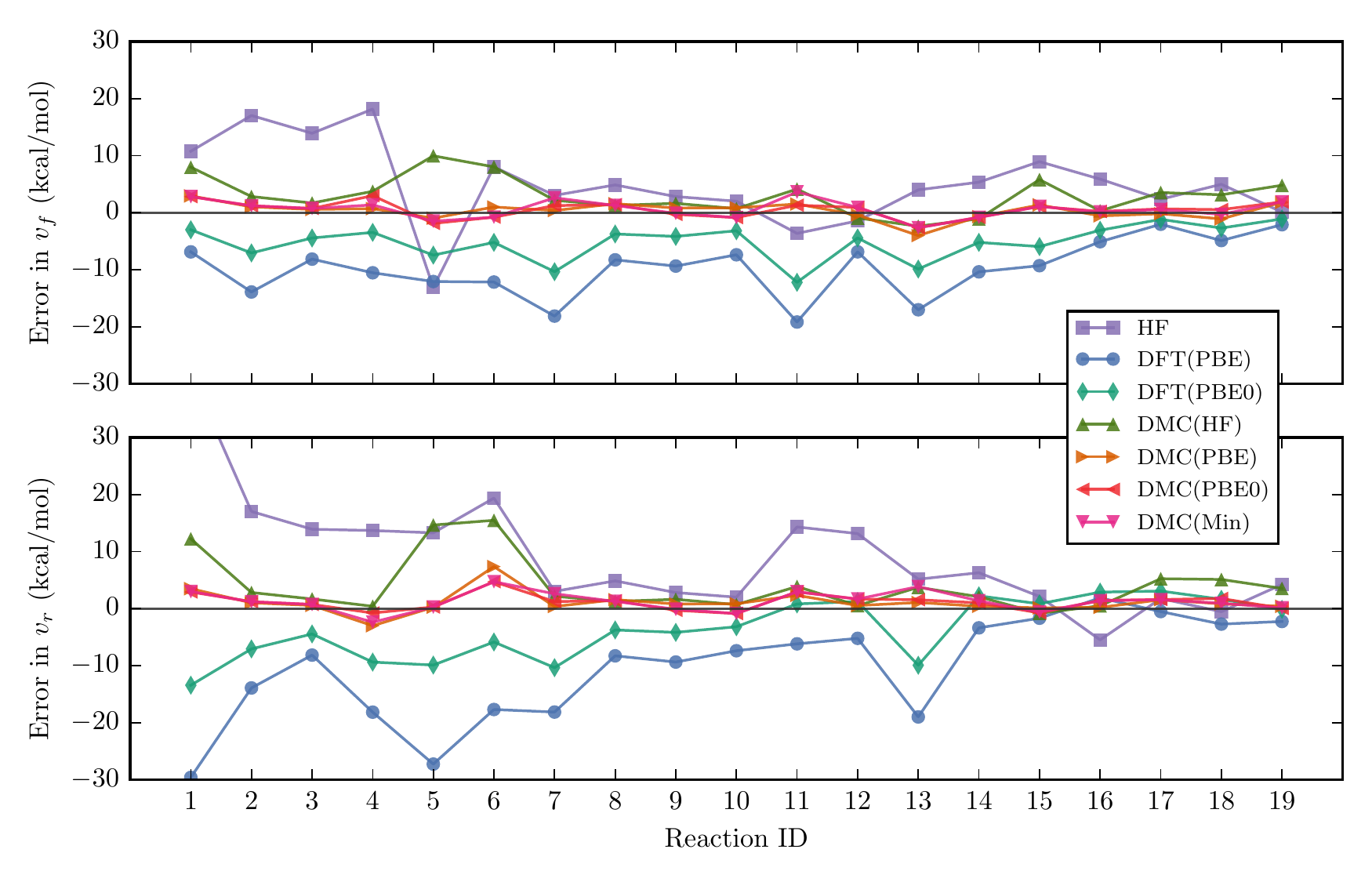}
\caption{\label{fig:verr_id}
The error of forward and reverse reaction barrier heights, denoted by $v_f$ and $v_r$, versus reaction ID as defined in Table~\ref{tab:id}.
The error bars are insignificant and therefore neglected. The value of the missing point, $v_r$ of HF for Reaction 1, is 41.14~kcal/mol.
}
\end{figure*}

After obtaining the energy for each system, we calculate the forward and reverse barrier heights, denoted by $v_f$ and $v_r$, of the reactions in Table~\ref{tab:id}.
The error between the calculated barrier heights and the experimental results \cite{zhao2005benchmark,peverati2014quest} are presented in Fig.~\ref{fig:verr_id}.
For reference, we also present the results of the DFT and HF calculations.
The DMC(PBE) and DMC(PBE0) are quite similar in behavior.
They tend to perform better than DMC(HF) whenever their results are significantly different.
Also, evident is that DMC tends to greatly improve the error over any of the methods we use as input for trial wave functions.
However, note that our DFT results are using HF pseudopotentials, so they will differ slightly from results including all electrons or using DFT- and DFT-functional-specific pseudopotentials.
The DMC results are also consistent in their accuracy, even when the accuracy of the method generating its trial wave function is unreliable.

Consistent with a previous study \cite{zhao2005benchmark}, HF tends to overestimate the barrier height, and the trend continues to hold for DMC(HF).
The notable exception is Reaction~5, which is far too low in HF, while it is unusually too high in DMC(HF).
This discrepancy can be understood by the fact that HF seems to overestimate the energy of \ce{F2} as can be seen from Fig.~\ref{fig:edmc}, while the fixed-node error of TS05 is particularly large.
This illustrates the general feature that the fixed-node errors tend not to correlate with the quality of the method generating the trial wave function, which is examined in the Discussion.

\begin{figure}
\includegraphics{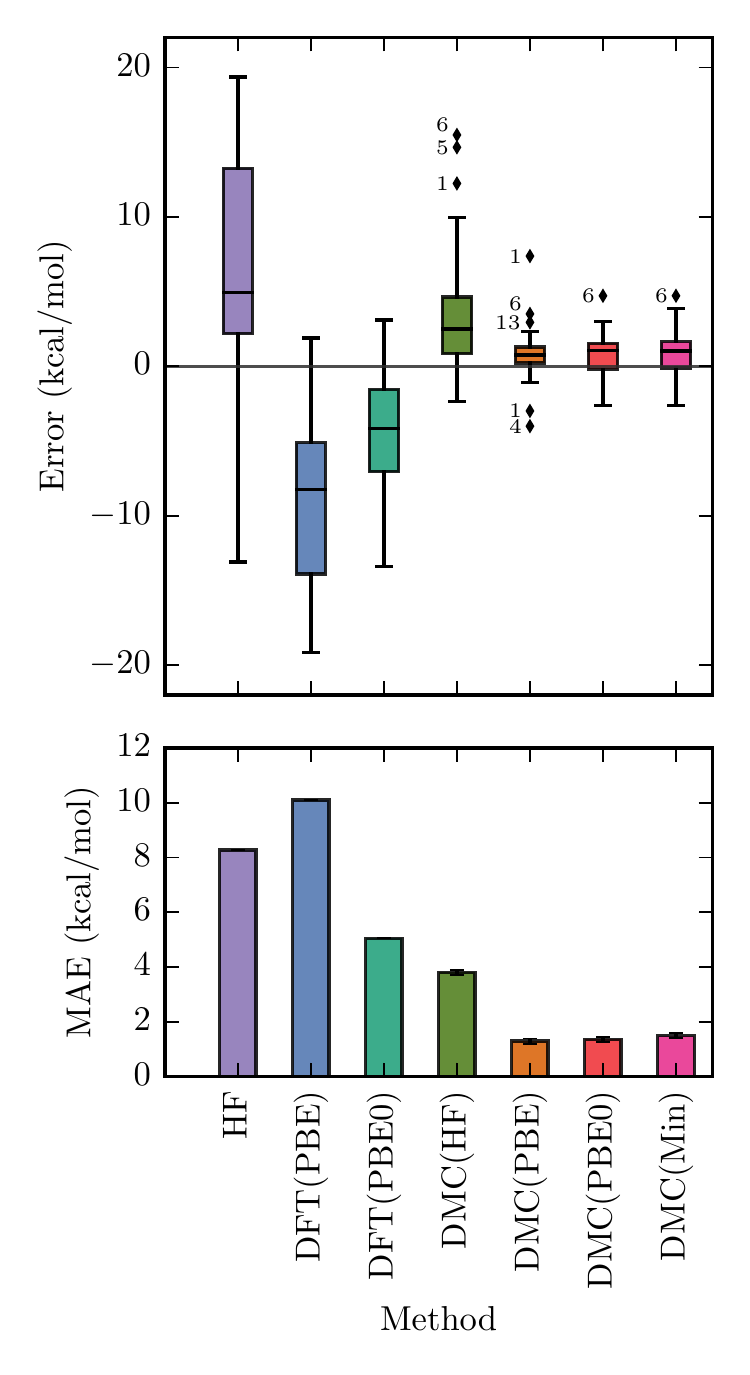}
\caption{\label{fig:box_bar}
(Top) The box-and-whisker plot of the barrier-height errors for each method.
Each box ends at first ($Q_1$) and third ($Q_3$) quantiles.
The horizontal line in each box represents the median.
The whiskers extend to the farthest points within 1.5 times the difference between $Q_1$ and $Q_3$.
Every point outside of this range is represented by a diamond and labeled by an ID number as defined in Table~\ref{tab:id}.
(Bottom) The bar chart of the mean absolute error for each method.
The DMC error bars denote statistical errors.}
\end{figure}

To summarize the results, the errors in $v_f$ and $v_r$ are combined into a single data set and
presented as a box-and-whisker plot \cite{chambers1983graphical} as shown in Fig.~\ref{fig:box_bar}~(Top).
The plot shows that DFT(PBE) tends to underestimate the barrier height, while HF does the opposite, as has been found before \cite{zhao2005benchmark,zhang2009doubly}.
On going to the DMC results, the spread in the barrier-height errors decreases dramatically, while the median is much closer to zero.
While DMC(PBE) method yields the median closest to zero, 5 out of the 19 results were outliers, suggesting this trial wave function is less reliable.

Finally, the mean absolute error for each method has been calculated and reported in a bar chart (Fig.~\ref{fig:box_bar})~(Bottom).
As discussed previously, the reactions being tested do not exhibit any significant difference between DMC(PBE0) and DMC(Min) because DMC(PBE0) tends to provide the lowest total energies.
Neglecting DMC(Min), the most accurate methods as shown are DMC(PBE) and DMC(PBE0) followed in order by DMC(HF), DFT(PBE0), HF, and DFT(PBE).
The differences in the overall accuracy for DMC(Min), DMC(PBE), and DMC(PBE0) are statistically indistinguishable by our calculations.
However, these three method clearly perform better than DMC(HF).
The error in DMC(PBE0) is close to the 1~kcal/mol accuracy necessary to predict chemical reaction rates.

\section{Discussion}
Analyzing trends in the fixed node error of the test set demonstrates the principle that the quality of the functional may have little to do with the quality of the DMC calculation utilizing the trial wave function it generates.
The error in the barrier heights themselves did not correlate reliably between the method producing the trial function and the final DMC result.
For instance, the HF error of $v_f$ in Reaction 5 is more negative, while DMC(HF) error is more positive compared to the average error.

We attempted to find some correlation with quantities computable in DFT and HF which may indicate the fixed node error may be large.
We computed several physical quantities within PBE0 and HF and checked to see if the differences between these results correlated with the nodal error, measured by the difference in energy between DMC(PBE0) and DMC(HF).
We found that the difference in energies computed by the PBE0 and HF did not correlated with the fixed-node error.
Also, the square difference in the electron densities between PBE0 and HF, $\int d^3r~(\rho_\text{PBE0}(\mathbf{r}) - \rho_\text{HF}(\mathbf{r}))^2$ did not correlated with the fixed-node error.
The total change in atomic charges, measured by $(\sum_I(c_{I,\text{PBE0}} - c_{I,\text{HF}})^2)^{1/2}$ where $c_{I,\text{PBE0}}$ is the number of electrons on ion $I$ as measured by PBE0, for example, also did not correlate with the fixed-node error.
Additionally, the difference in the barrier heights between DMC(PBE0) and DMC(HF) did not correlate with the difference in barrier heights between PBE0 and HF.

\begin{figure}
\includegraphics{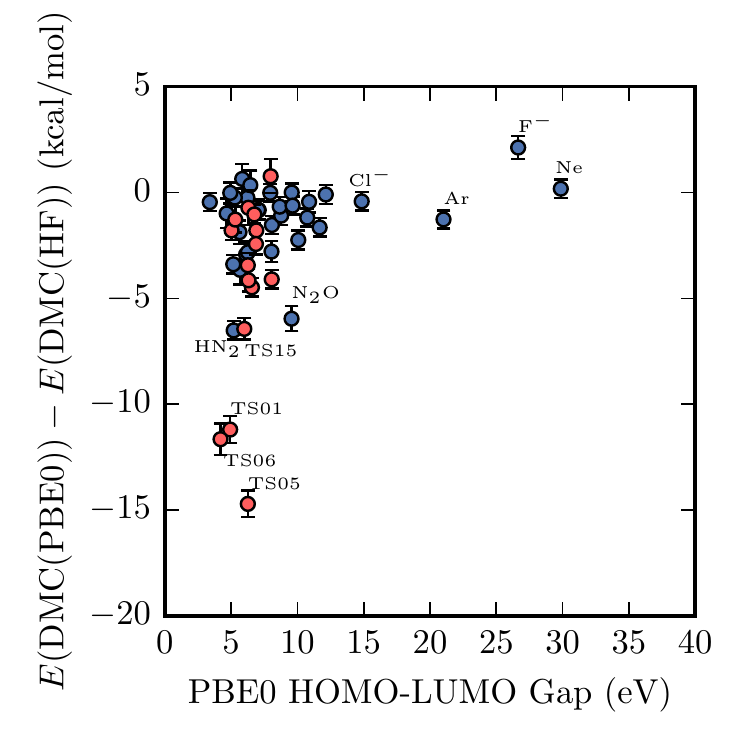}
\caption{\label{fig:edmc_gap_pbe0_min} Approximate nodal error, measured by the energy difference between DMC(PBE0) and DMC(HF) plotted against theHOMO-LUMO gap, computed by PBE0.
The transition states are shown in coral.
The other systems are shown in blue.}
\end{figure}

We did find some correlation between the energy  HOMO-LUMO gap and the fixed-node error.
Fig.~\ref{fig:edmc_gap_pbe0_min} plots out the HOMO-LUMO gap computed by PBE0 compared to the energy difference between DMC(PBE0) and DMC(HF).
This plot illustrates how the worst fixed-node errors tend to occur in transition states, and these transition states tend to have lower energy gaps.
This is consistent with previous work on reaction barriers~\cite{fracchia_barrier_2013}, which found that transition states tend to have multiconfigurational character, due to the stretching of bonds which often occurs in transition states.

\section{Conclusion}
We have found that fixed node diffusion Monte Carlo (DMC) with a single Slater determinant can obtain near-chemical accuracy for a benchmark set of 19 chemical reactions.
Using this set, we performed statistical analysis to investigate trends in the nodal error.
Of the functionals we surveyed, PBE0 provides the lowest energy nodal surfaces for almost all molecules.
The size of the nodal errors in DMC are uncorrelated with the error of the DFT functional used to produce the trial wave function but do tend to be larger for transition states with small HOMO-LUMO gaps.
However, the HOMO-LUMO gap does not appear to completely determine the nodal error.
Since the version of the algorithm we used scales very well with system size, $\mathcal{O}(N_e^{3-4})$, it is applicable even to larger systems.
From these results, it appears that DMC could be a viable route to performing high accuracy calculations on barrier heights for many chemical systems.

\section{}
This material is based upon work supported by the U.S. Department of Energy, Office of Science, Office of Advanced Scientific Computing Research, Scientific Discovery through Advanced Computing (SciDAC) program under Award Number FG02-12ER46875.
B.B. was supported by the National Science Foundation Graduate Research Fellowship Program.
Computational resources were provided by the Illinois Campus Cluster program.

\bibliography{paper}
\end{document}